\pdfoutput=1
\documentclass[a4paper, 10pt, conference]{ieeeconf}  % Comment this line out if you need a4paper

\IEEEoverridecommandlockouts                              % This command is only needed if 

\usepackage{cite}
\usepackage{mathtools}
\usepackage{amsmath,amssymb,amsfonts}
\usepackage{algorithmic}
\usepackage{graphicx}
\usepackage{textcomp}
\usepackage{xcolor}
\usepackage{algorithmic}
\usepackage[ruled,vlined]{algorithm2e}

\usepackage{hyperref}

\title{\LARGE \bf
Deep DIH : Statistically Inferred Reconstruction of Digital In-Line Holography by Deep Learning
}

\author{
Huayu Li$^{1}$, Xiwen Chen$^{1}$, Haiyu Wu$^{1}$, Zaoyi Chi$^{1}$, Christopher Mann$^{2}$, and Abolfazl Razi$^{1}$
\thanks{$^{1}$School of Informatics, Computing and Cyber Systems, Northern Arizona University}
\thanks{$^{2}$Department of Physics and Astronomy Physical Sciences Building, Northern Arizona University}
}

\begin{document}

\maketitle
\thispagestyle{empty}
\pagestyle{empty}

%%%%%%%%%%%%%%%%%%%%%%%%%%%%%%%%%%%%%%%%%%%%%%%%%%%%%%%%%%%%%%%%%%%%%%%%%%%%%%%%
\begin{abstract}
Digital in-line holography is commonly used to reconstruct 3D images from 2D holograms of microscopic objects. One of the technical challenges that arises in the signal processing stage is that of the elimination of the twin image originating from the phase-conjugate wavefront. The twin image removal is typically formulated as a non-linear inverse problem due to the irreversible scattering process when generating the hologram. Recently, end-to-end deep learning methods have been utilized to reconstruct the object wavefront (as a surrogate for the 3D structure of the object) directly from a single-shot in-line digital hologram. However, massive data pairs are required to train deep learning models for acceptable reconstruction precision. In contrast to typical image processing problems, well-curated datasets for in-line digital holography do not exist. Also, the trained model is highly influenced by the morphological properties of the object and hence can vary for different applications. As a result, data collection can be prohibitively lengthy and cumbersome and currently represents a major obstacle in utilizing deep learning for digital holography. In this paper, we proposed a novel implementation of autoencoder-based deep learning architecture for single-shot hologram reconstruction based solely on the current sample and without the need for massive datasets to train the model. The simulations results demonstrate the superior performance of the proposed method compared to the state of the art single-shot compressive digital in-line hologram reconstruction method.
\end{abstract}

%%%%%%%%%%%%%%%%%%%%%%%%%%  body  %%%%%%%%%%%%%%%%%%%%%%%%%%
\section{Introduction}
Digital Holography is a powerful imaging technique which is used to record the information of the three-dimensional (3D) surface of an object from a two dimensional (2D) image captured by visual sensors. It is mainly used for the investigation of micro-scaled as well as nano-scaled objects, and is used in wide range of different applications areas such as chemistry~\cite{matsui2019chemical}, biomedical microscopy \cite{xu2001digital}, nano-material fabrication~\cite{zheng2015metasurface,ramanujam2005optical}, and nano-security~\cite{van2000synergistic}. 

\begin{figure}[htbp]
\centering\includegraphics[width=0.9\columnwidth]{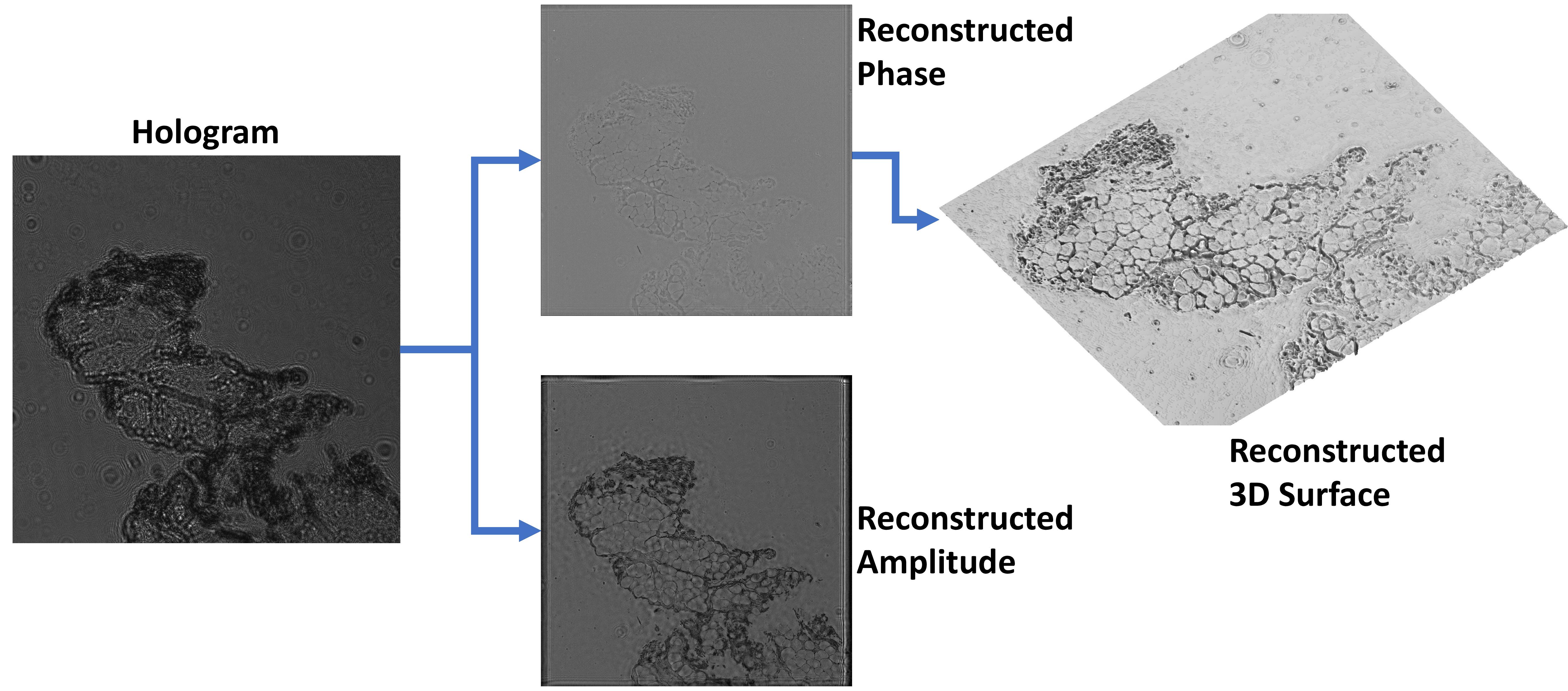}
\caption{The amplitude and phase of the wavefront can be extracted from the recorded hologram using numerical methods. The phase information represents the surface depth or the thickness of the object, that can be used to reconstruct the 3D view of the object. This figure shows that how digital holography record the 3D information of an object into 2D form.}
\label{fig1}
\end{figure}

Digital holography can be used in several different modalities, including that of in-line digital holography transmission imaging for mostly transparent objects \cite{cuche1999simultaneous}. The sample modulates the wavefront phase of the emitted linearly-polarized laser beam. The 3D structure of the object can be easily reconstructed from the recovered phase information, as shown in Fig.\ref{fig1}.

Regardless of the object type, there are two different implementation approaches for digital holography, off-axis holography~\cite{leith1963wavefront}, and in-line holography~\cite{gabor1948new}. In Off-axis holography, the laser beam is split into two waves, the reference wave denoted by $R$ and object wave denoted by $O$, where only the latter passes through the object. The two waves are combined with a small relative incidence angle $\theta$ at the exit of the interferometer to create the hologram intensity as $I_H(x,y)=|R|^2+|O|^2+R^*O+RO^*$, where $X^*$ denotes the complex conjugate of $X$. The relative angle causes the real images and twin images to formed in separable locations in Fourier space. This spatial separation facilitates easier phase recovery through filtering in the Fourier domain. However, this method faces practical implementation problems as it requires an accurate synchronization between the reference and object waves that become prohibitively hard for nano-scaled imaging. An accurate characterization of the reference wave based on the Fresnel–Kirchhoff integral is also required for the numerical phase reconstruction \cite{cuche1999simultaneous,schnars2002digital}. Digital in-line holography (DIH) uses only a single laser beam with numerical reconstruction by the angular spectrum algorithm for phase retrieval. Other advantages of DIH, include the elimination of the need for objective lenses, the simplicity of sample preparation with no need for sectioning and staining, as well as its high-speed imaging capabilities~\cite{xu2001digital}. 

\begin{figure}[htbp]
\begin{center}
\centerline{\includegraphics[width=0.7\columnwidth]{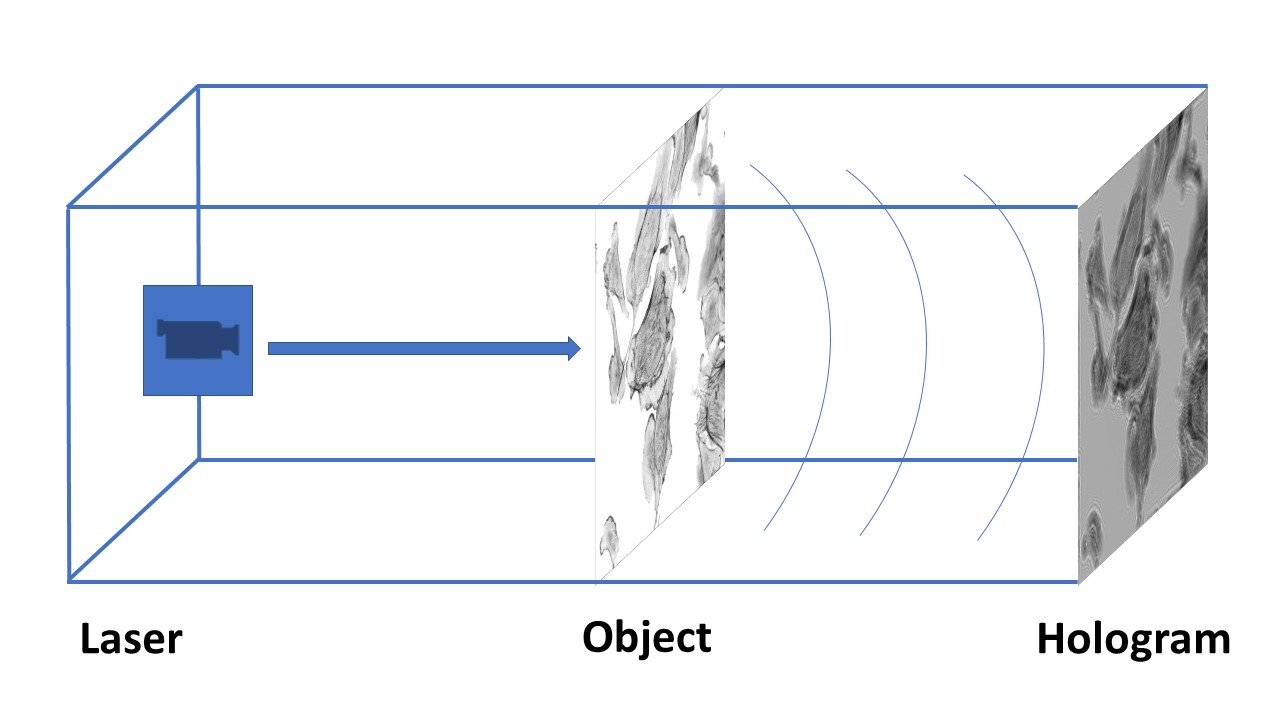}}
\caption{The twin-image issue: the scattered object wave interferes with the unscattered reference wave in the inline holography.}
\label{fig2}
\end{center}
\end{figure}

To further explore the physical model behind the concept of twin image removal, we investigate the process of inline holography, as shown in Fig.\ref{fig2}. Suppose that we have an object field $\rho(x,y)$ and the propagation transfer function  function $h(x,y)$, the scattered wave $O(x,y)$ can be described as~\cite{zhang2018twin}:
\begin{equation}
    O(x,y) = \int \int _{x_i,y_i\in \Sigma}\rho(x_{i},y_{i})h(x-x_{i},y-y_{i})dx_{i}dy_{i}
\end{equation}
where $\Sigma$ represents a aperture window. The transmittance function $h(x,y)$ depends on the light wavelength $\lambda$ and the propagation distance $z$ between the image plane and the hologram. The transfer function in frequency domain is:
\begin{equation}
    H(f_x,f_y)=\text{exp}(ikz\sqrt{1-(\lambda f_{x})^{2}-(\lambda f_{y})^{2}})
\end{equation}
where $k=2\pi/\lambda$ is the wave number. In addition to the diffracted wave $O(x,y)$, there exists a non-scattered reference wave $R(x,y)$. The Hologram $I_H(x,y)$ records the intensity of the mixed waves captured by the light sensors and can be expressed as:
\begin{equation}
\begin{aligned}
    I_H&=|O+R|^{2}=O^{*}R+OR^{*}+|O|^{2}+|R|^{2}\\&=U(x,y)+U^*(x,y)+|O|^{2}+|R|^{2}
\end{aligned}
\end{equation} 
where we define $U(x,y) = O∗R$ for notation convenience. The captured hologram includes the object field $O(x,y)$ and its conjugation $O∗(x,y)$, respectively, representing the virtual and real images~\cite{cuche1999simultaneous}. This phenomenon leads to the twin image problem present during the reconstruction. As one focuses on one of the holographic terms, the out of focus conjugate smears the reconstructed image.

Noting that the unscattered field ($|R|^2$) can be assumed one with the loss of generality and can be removed from the hologram. Also, the term $|O|^2$ can be regarded as the noise term $n(x,y)$. Therefore, the problem of reconstructing the object field boils down to removing the twin image \cite{zhang2018twin}, which has been the center of attention in many prior works~\cite{allen2001phase,xu2001digital,waller2010transport,denis2005twin}. If we define transformation $T:\rho(x,y)\mapsto U(x,y)$. 
Therefore, the image reconstruction can be recast as the following standard inverse problem:
\begin{equation} \label{eq:inv3}
I_H(x,y)=2Re\big[T\big(\rho(x,y)\big)\big]+n(x,y)    
\end{equation}
Both $U(x,y)$ and its conjugation $U^{*}(x,y)$ are interchangeably consistent with the solution of Equation.~\ref{eq:inv3} which could both be the solution to this problem, the reconstruction of the digital in-line holography is typically under-determined. Also, standard inverse problems may not be utilized to solve Equation.~\ref{eq:inv3}, as it includes the non-linear transformation and the symmetric diffracting which towards the opposite direction. 

There exist several means for solving the twin image problem. Recording a collection of holograms at different propagation distances and reconstructing the object field by the Transport of Intensity (TIE) method has yield promising results~\cite{tong1991energy,barton1991removing}. Most conventional phase retrieval methods use the following TIE imaging equation to recover the phase term $\phi(x,y)$~\cite{teague1983deterministic,waller2010transport,gureyev1995partially}: 
\begin{equation}
\label{eq:tie1}
    \frac{\partial I(x,y)}{\partial z} = -\frac{\lambda}{2\pi}\nabla(I(x,y)\nabla \phi(x,y)) 
\end{equation}

where $I(x,y)$ is the hologram intensity, $\lambda$ is the wavelength, and $\nabla$ is the gradient operator in the lateral dimensions $(x,y)$~\cite{waller2010transport}. When the intensity is constant (or normalized), the following simplified equations can be used to recover $\phi(x,y)$~\cite{waller2010transport,zuo2013direct,allen2001phase}:
\begin{equation}  
\label{eq:tie2}
    \frac{2\pi}{\lambda I} \frac{\partial I(x,y)}{\partial z}= \nabla^2 \phi(x,y)
\end{equation}
Since then, several extensions to the TIE method are proposed in the literature to extend it for different applications including volume holography \cite{waller2010volume}, and holographic x-ray imaging \cite{krenkel2013transport}. One technical difficulty in solving Equation.~\ref{eq:tie1} and Equation.~\ref{eq:tie2} is the need for multiple imaging at fine-tuned distances from the focal plane (i.e, $\Delta z$, $2\Delta z$, $\dots$) to precisely quantify the gradient term ${\partial I(x,y)}/{\partial z}$ using least square method \cite{soto2007improved}, hybrid linearization method \cite{gureyev2003composite,gureyev2006linear}, and iterative methods \cite{allen2001phase}. 

Therefore, developing methods that can recover phase information from only one measurement has obvious practical advantages. Phase retrieval (PR) is one of the most commonly used numerical approaches which perform double-side constraint iteration with a specific support region.  Mathematically, the in-line hologram provides an undesirable component that can be traced to the loss of phase information. PR permits the separation of real-object distribution from the twin-image interference. Gerchberg-Saxton (GS) algorithm ~\cite{gerchberg1972practical,fienup1982phase,zalevsky1996gerchberg,denis2005twin} and Hybrid input-output (HIO) algorithm~\cite{bauschke2002phase,fienup1978reconstruction} perform iterative phase retrieval followed the below steps:
\begin{itemize}
    \item \textbf{Step 1:} Let $\rho^{(n)}$ be a trial scattering density in the $n^{th}$ iteration cycle.
    \item \textbf{Step 2:} Let $\rho^{'(n)}$ be a density obtained from $\rho^{(n)}$ by Fourier transform.
    \item \textbf{Step 3:} Replacing all Fourier amplitudes by the experimentally observed amplitudes, and applying inverse Fourier transform.
    \item \textbf{Step 4:} Imposing constraints to the object plane in the support region.
\end{itemize}
the support region is usually designed based on a known prior. In GS algorithm, the object plane $\rho^{n}$ in the support region $\gamma$ are constraint as:
\begin{equation}
    \rho^{n+1}= \left\{
\begin{array}{ll}
        0       &       \mbox{$\rho^{n} \in \gamma$};\\
        \rho^{'n}       &      \mbox{$\rho^{n} \notin \gamma$};
\end{array} \right.
\end{equation}

while the HIO algorithm deploys a relaxing factors $\beta$ to reduce the probability of stagnation that contains feedback information concerning previous iterations as:
\begin{equation}
    \rho^{n+1}= \left\{
\begin{array}{ll}
        \rho^{n}-\beta \rho^{'n} &      \mbox{$\rho^{n} \in \gamma$};\\
        \rho^{'n}       &      \mbox{$\rho^{n} \notin \gamma$};
\end{array} \right.
\end{equation}
Although PR shows excellent performance on the object reconstruction. Due to the double-side constraint iteration with a specific support region, the reconstruction area is under a severe limitation.

Recently, deep learning based approaches~\cite{horisaki2018deep,rivenson2018phase,gan2017holography} were proposed for end-to-end digital hologram reconstruction and proven effective by utilizing the outstanding learning capability of deep convolutional neural networks (CNNs). As a universal approximator, CNNs are widely used in solving inverse problems in the field of computer vision. The general workflow of the deep learning method is first training a CNN on labeled data pairs (holograms, and twin image free phase and amplitude), then using the well-trained CNNs to predict the unlabeled data. Deep learning based methods are typically data-driven approaches that massive data pairs are needed for training the CNN. In most natural image processing tasks, massive data pairs are easily accessible. Unfortunately, digital holography is usually deployed in biomedical imaging that getting large amounts of data is costly since both capturing holograms and generating the corresponding ground truth is pretty difficult. Meanwhile, the CNNs are regarded as black boxes when the training and inferring steps are invisible and unexplainable. That means when using a well trained CNN to reconstruct the hologram, it is impossible to deal with the upcoming problems if the reconstruction is not correct. 

In~\cite{zhang2018twin}, a compressive sensing (CS) approach to reconstruct a twin image free hologram was proposed. The CS method is able to remove the twin image with single-shot hologram and does not need massive training pairs. As a physics-driven method, the CS method lies on the sparsity difference between the reconstructed object and the twin image that filters out the diffuse conjugated signal by imposing sparsity constraints on the object plane. Total variation (TV) norm is suitable for removing twin image since the in-focus object has sharp edges while the out-of-focus twin image is diffuse. A two-step iterative shrinkage/thresholding (TwIST) algorithm is used in~\cite{zhang2018twin} to address the twin image removal problem by minimizing an objective function formed by Mean Square Error (MSE) and TV norm:
\begin{equation}
    \hat{\rho} = \arg \underset{\rho}{min} \left \{ \frac{1}{2} ||H-T(\rho)||^2_2+\tau||\rho||_{tv} \right \}
\end{equation}
where $\tau$ is the relative weight between the TV norm $||\rho||_{tv}=\sum_{i}\sqrt{|\Delta^x_i\rho|^2+|\Delta^y_i\rho|^2}$ and the MSE term. The $\Delta^x_i$ and $\Delta^y_i$ refers to the horizontal and vertical first-order gradients. Based on the idea proposed in~\cite{langehanenberg2008autofocusing}, the reconstruction with a more dense edge matrix commonly suffers a more out-of-focus twin image as well as has a larger TV norm. The CS method has been proven more effective than PR that can reconstruct a more clear and twin image free hologram. It still has a couple of problems. Deploying TV norm to remove the twin image should make a trade-off on the relative weight $\tau$. Since large values of $\tau$ lead to blur the reconstruction and small values of $\tau$ have a weak effeteness on twin image removal. Also, imposing sparsity constraints on an image restoration problem leads to edge distortion.

In this paper, a novel deep learning implementation based on fitting an untrained auto-encoders to the possible solutions of a single captured hologram through minimizing a physics-driven object function. This method performs noise reduction and twin image removal simultaneously and does not require massive data to train the model. In the presented manner, we do not suppress or remove the twin image in the reconstruction. Instead, we directly fit the CNNs to search the possible intensity and phase of the target 3D object consistent with the captured hologram. We show that neural networks equipped with convolutional layers naturally tend to produce a more transparent result. Experimental results prove the feasibility and the superior performance of the proposed method over the existing CS methods.

\section{Deep Learning Scheme}
\label{deep}
\begin{figure}[htbp]
\begin{center}
\centerline{\includegraphics[width=0.9\columnwidth]{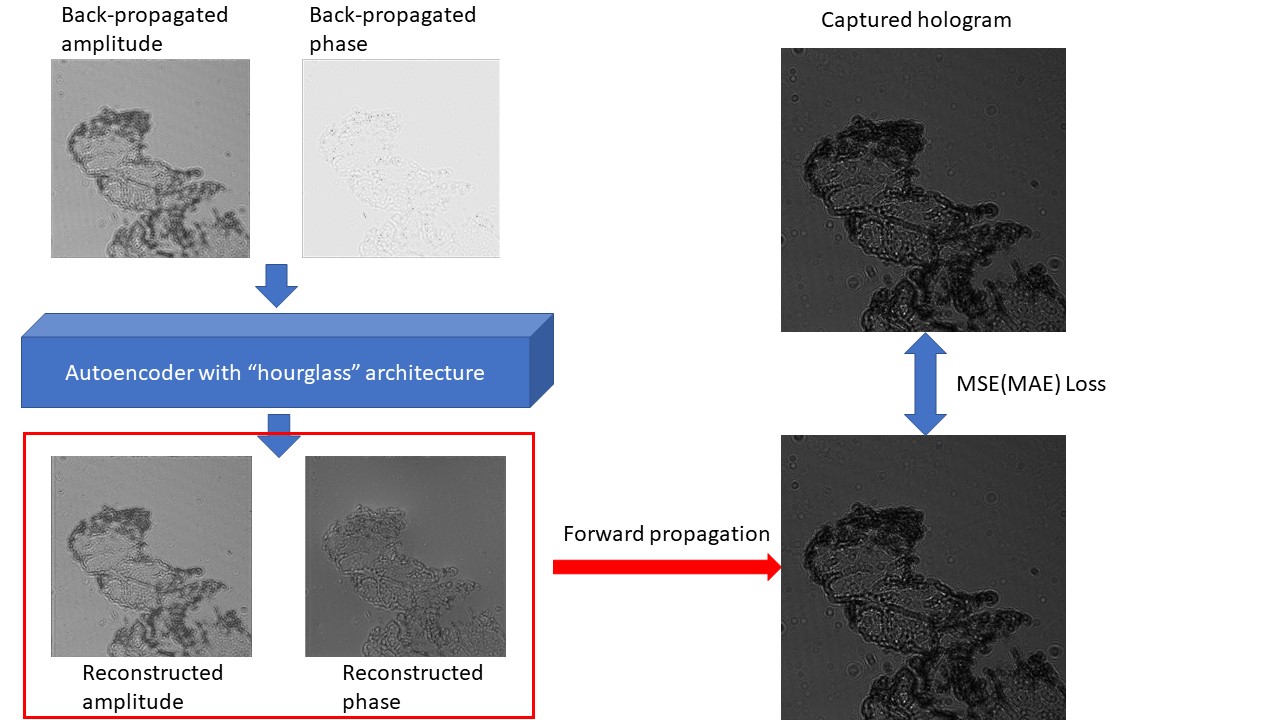}}
\caption{This figure shows the learning procedure of the proposed method. After feeding a fixed input into the network, the network generates a reconstructed result. The reconstructed result will be propagated to the hologram plane by the transmission depending on the optical parameters. The network updates its weights by minimizing the pixel error between the forward-propagated results and captured hologram.}
\label{fig3}
\end{center}
\end{figure}

A deep network with encoder-decoder architecture which is also called auto-encoder maps a high dimensional input $x$ into low dimensional latent code $z=f_{encode}(x)$ and reconstruct a high dimensional output $\hat{x}=f_{decode}(z)$ from the latent code. The common formulation used in supervised image restoration is to minimize the error between the output $\hat{x}$ and the ground truth $y$. In~\cite{ulyanov2018deep}, an unsupervised blind image restoration called Deep Image Prior (DIP) has proven that fit a randomly initialized CNN to a single corrupted image is able to recover the clean image since the CNNs could naturally learn the uncorrupted and realistic part. Inspire by DIP, we consider using the same scheme with DIP to remove the twin image in the reconstructed object plane. But there arises another problem that there is a high coupling between the virtual and real object plane in both spatial and frequency domain, and the CNNs will generate an output with the twin image. As mentioned in~\cite{zhang2018twin}, the twin image term is denser than the object term. Here we investigate a novel learning procedure that using the physical model in the objective function in the training process, as shown in Fig.~\ref{fig3}. Assume there is an autoencoder with random initiated weights $w$, the output reconstruction $\rho$ can be expressed as $\rho = f(x,w)$, where $x$ is a fixed input. And the objective function could be formulated as:
\begin{equation}
    w = \mathop{\arg\min}_{w} \ \ \| H-T(f(H,w))\|_{2}^{2}
\end{equation}
where we want to propagate the reconstruction to the hologram plane with the transmission $T$ and minimize the error between the captured hologram and the forward-propagated results. When minimizing the object function, the network actually performs searching the possible results from parameter space. Through the experiments we conduct, which will be shown later in this paper at Section.~\ref{sim}, the network tends first to generate the primary instance, which is the rough shape of the reconstructed object. Then the network gradually recovers the details from different levels of the object. This phenomenon usually causes the network applied for other image recovery tasks such as denoising and super-resolution overfit the degraded term in the corrupted image. However, in the case of hologram reconstruction, both the twin image and clean object could be the solution of the non-linear inverse problem. Therefore, after generating the main body of the object, the network continues to generate the real details instead of the twin image.  

\section{Implementation of Auto-encoder}
We use the wavelet transform as the downsampling method as an alternative of pooling or strided convolution. According to the previous work~\cite{rivenson2016sparsity}, using a wavelet transform could impose sparsity on the reconstruction object plane. Therefore, we take Haar wavelet and its inverse transform as the downsampling and upsampling method in our network. The Haar wavelet decomposes the input image or feature map into four sub-band by four convolutional filters (one low pass filter $f_{LL}$, three high pass filters $f_{LH}$, $f_{HL}$, and $f_{HH}$). The four filters are defined as: $f_{LL}$ = $ \left[ \begin{smallmatrix} 1 & 1 \\ 1 & 1 \end{smallmatrix}  \right] $, $f_{LH}$ = $ \left[ \begin{smallmatrix} -1 & -1 \\ 1 & 1 \end{smallmatrix}  \right] $, $f_{HL}$ = $ \left[ \begin{smallmatrix} -1 & 1 \\ -1 & 1 \end{smallmatrix}  \right] $, and $f_{HH}$ = $ \left[ \begin{smallmatrix} 1 & -1 \\ -1 & 1 \end{smallmatrix}  \right] $. The four sub-bands are obtained by convolution operation as $x_{LL}=(f_{LL}\otimes x)$, $x_{LH}=(f_{LH}\otimes x)$, $x_{HL}=(f_{HL}\otimes x)$, and $x_{HH}=(f_{HH}\otimes x)$, where $\otimes$ refers to convolution operator. The inverse transform of Haar wavelet in $(x,y)-th$ pixels can be written as:
\begin{flalign}
\begin{split}
&x(2i-1,2j-1) =\frac{1}{4}(x_{LL}(i,j)-x_{LH}(i,j)-x_{HL}(i,j)\\&+x_{HH}(i,j))\\
&x(2i-1,2j) =\frac{1}{4}(x_{LL}(i,j)-x_{LH}(i,j)+x_{HL}(i,j)\\&-x_{HH}(i,j))\\
&x(2i,2j-1) =\frac{1}{4}(x_{LL}(i,j)+x_{LH}(i,j)-x_{HL}(i,j)\\&-x_{HH}(i,j))\\
&x(2i,2j) =\frac{1}{4}(x_{LL}(i,j)+x_{LH}(i,j)+x_{HL}(i,j)\\&+x_{HH}(i,j))\\
\end{split}
\end{flalign}

\begin{figure*}[htbp]
\begin{center}
\centerline{\includegraphics[width=1.5\columnwidth]{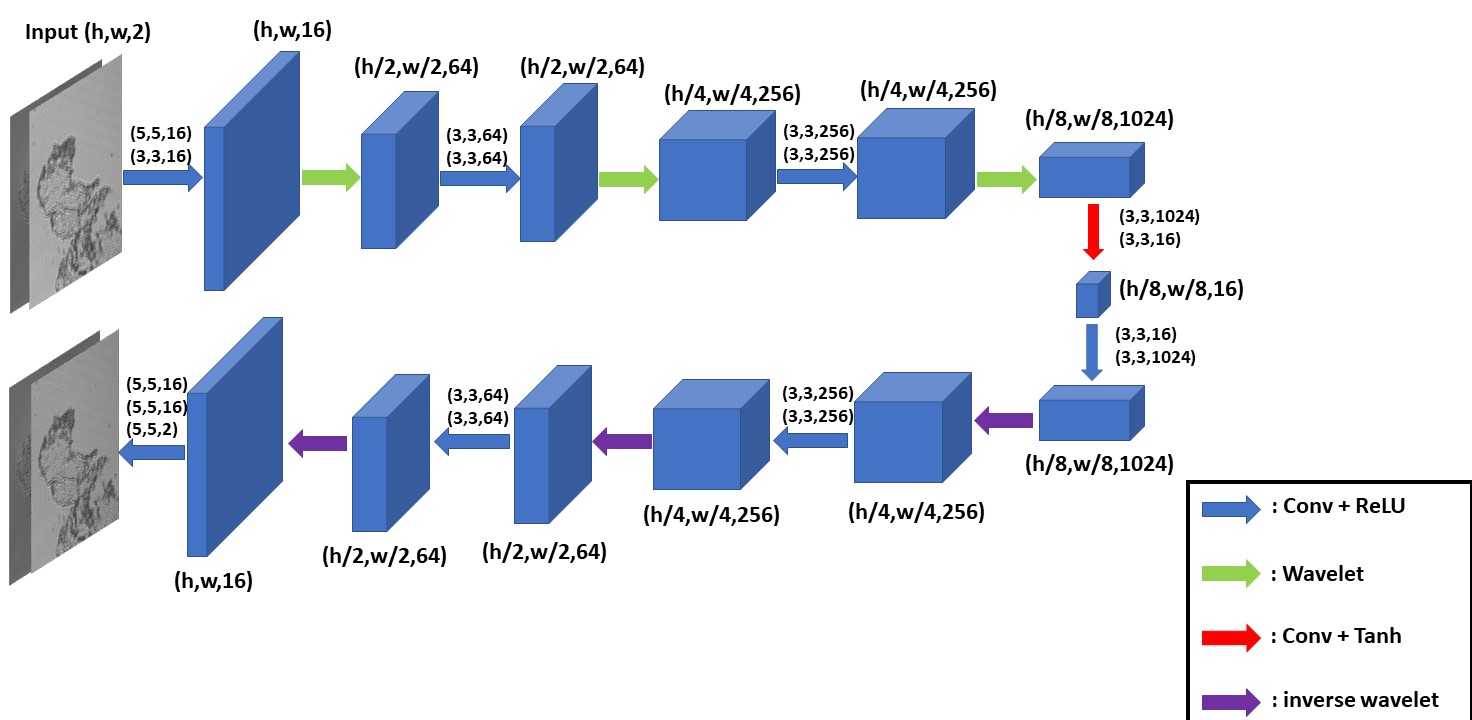}}
\caption{This figure shows the deep convolutional autoencoder with “hourglass” architecture used in this paper. We deploy Batch Normalization~\cite{ioffe2015batch} after each convolution layer except the last three layers to stabilize the training steps.}
\label{fig4}
\end{center}
\end{figure*}

We first build a Auto-encoder with "Hourglass" architecture, as shown in Fig~\ref{fig4}. The encoder $f_{e}(\bar{\rho})$ maps the fixed network input into lower-dimensional manifold, and the decoder $f_{d}(f_{e}(\bar{\rho}))$ recover the object we want from the latent code. It is noticeable that during our experiment, we found that if skip-connection is used in the CNNs, the network will identically map the input to the output instead of searching a possible result. 

\section{Simulation Results}
\label{sim}
In this section, several comparison experiments with the CS method used in \cite{zhang2018twin} are conducted on several simulated holograms to verify the feasibility of the presented methodology. We implement our model using the PyTorch Framework~\cite{paszke2019pytorch} in a GPU workstation with an NVIDIA Quadro RTX5000 graphics card. Adam optimizer~\cite{kingma2014adam} is adopted and set with a fixed learning rate at 0.0005. We train the network for 1500 to 3500 epochs for different holograms. For CS method, We set the relative weight of TV norm between $0.01$ to $0.1$ based on different holograms, as well as training iteration between $150$ to $350$. 

Three metrics are used to evaluate the reconstruction quality. The mean squared error (MSE) measures the average of the squares of the pixel-wise errors between ground truth image and reconstructed image, which is defined as:
\begin{equation}
    MSE(x,y) = \frac{1}{MN}\sqrt{\sum^{N,M}_{i,j} (x_{i,j}-y_{i,j})^{2}}
\end{equation}
Peak signal-to-noise ratio (PSNR) is an engineering term for the ratio between the maximum possible power of a signal and the power of corrupting noise that affects the fidelity of its representation. PSNR is most easily defined via the MSE that can be expressed as:
\begin{equation}
    PSNR(x,y) = 10 \log_{10}\frac{R^{2}}{MSE(x,y)}
\end{equation}
The Structural Similarity Index (SSIM)~\cite{wang2004image} is a perceptual metric that quantifies image quality degradation* caused by processing such as data compression or by losses in data transmission. The SSIM has been proven to be more consistent with the human visual system when compared to PSNR and MSE that the SSIM quantifies the changes in structural information by inspecting the relationship among the image contrast, luminance, and structural components. The SSIM between two images is given by:
\begin{equation}
   SSIM(x,y) = \frac{(2\mu_{x}\mu_{y}+C_{1})(2\sigma_{xy}+C_{2})}{(\mu_{x}^2+\mu_{y}^2+C_{1})(\sigma_{x}^2+\sigma_{y}^2+C_{2})}
\end{equation}
where $\mu_{x}$, $\mu_{y}$, $\sigma_{x}$, $\sigma_{y}$, and $\sigma_{xy}$ are the local means, standard deviations, and cross-covariance for images $x, y$. $C_1$ and $C_2$ are two variables to stabilize the division with a weak denominator.

\begin{figure}[htbp]
\begin{center}
\centerline{\includegraphics[width=0.9\columnwidth]{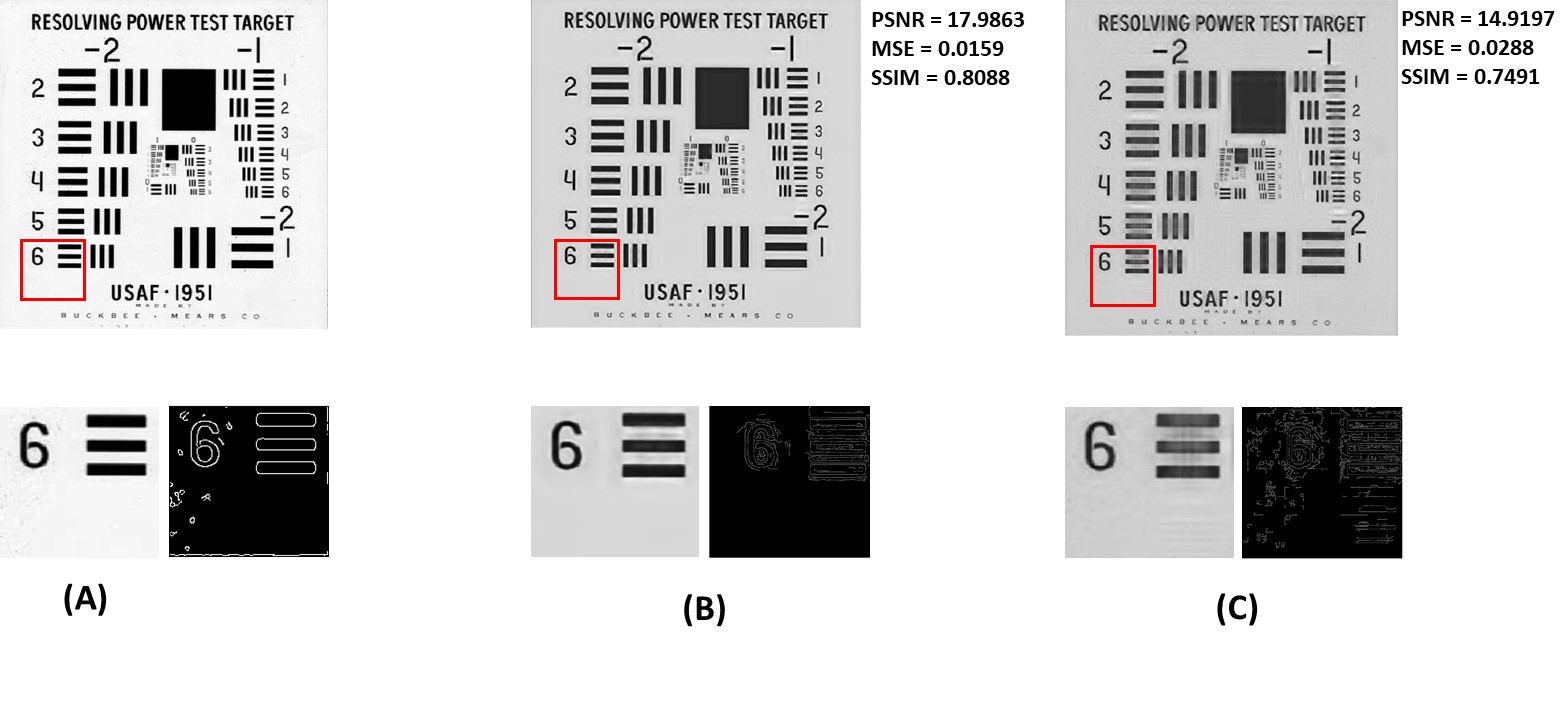}}
\caption{The reconstruction intensity of USAF resolution chart, the enlarged area from reconstruction ,and the edge matrix obtained by Canny edge. (A) Ground Truth. (B) Our method. (C) CS method. Although the proposed method does not use TV norm as prior to remove the twin image, it still reconstruct a more clear image with sparser edge matrix compare to the CS method~\cite{zhang2018twin}.}
\label{fig6}
\end{center}
\end{figure}

Fig.~\ref{fig6} compares the amplitude reconstruction of the proposed method with the CS method on the simulated USAF resolution chart. The image is resized into $1000\times1000$ pixels. The illumination light is set with a wavelength at 532 $\mu m$, and a complementary metal-oxide-semiconductor(CMOS) sensor is set with a pixel size of 4 $\mu m$. The distance between objects and the sensor is 1.2 $cm$. The proposed method produces a higher quality reconstructed image, which is more similar to the ground truth. Also, a Canny edge detector is used to extract the edge matrix of the enlarged area in these two reconstruction results and ground truth images. The edge matrix shows that the presented method has a better denoising capacity than the CS method, even without any hand-craft prior such as the TV norm. 

\begin{figure}[htbp]
\begin{center}
\centerline{\includegraphics[width=0.9\columnwidth]{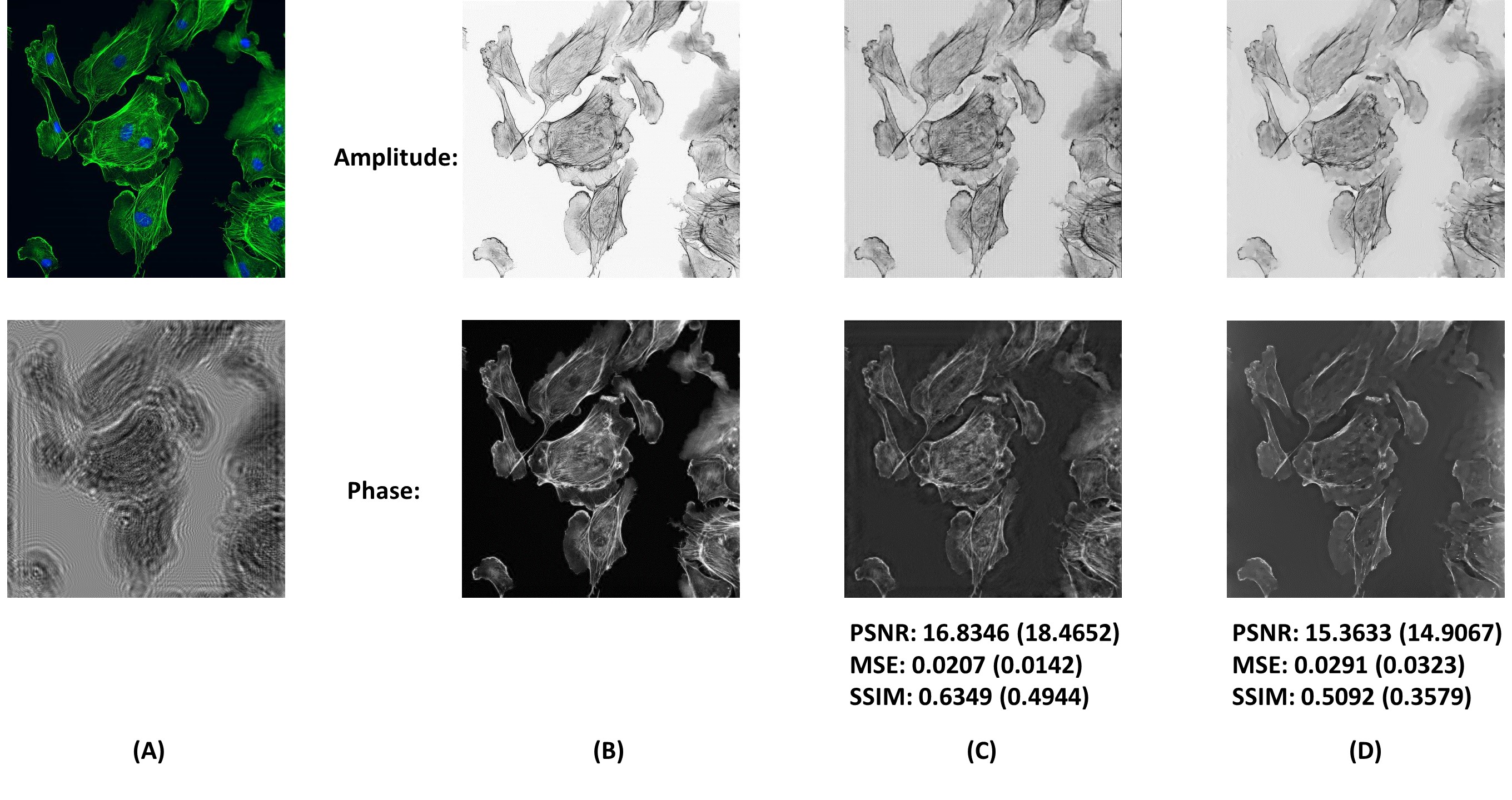}}
\caption{Cell image reconstructions and the evaluating metrics for amplitude (phase): (A) The RGB image and simulated hologram. (B) The reference amplitude and phase. (C) Our method. (D) CS. Here we can see that our method reconstruct a image with more clear detailed texture both for amplitude and phase that let the result has a higher SSIM and PSNR with the ground truth image.}
\label{fig7}
\end{center}
\end{figure}

\begin{figure}[htbp]
\begin{center}
\centerline{\includegraphics[width=0.9\columnwidth]{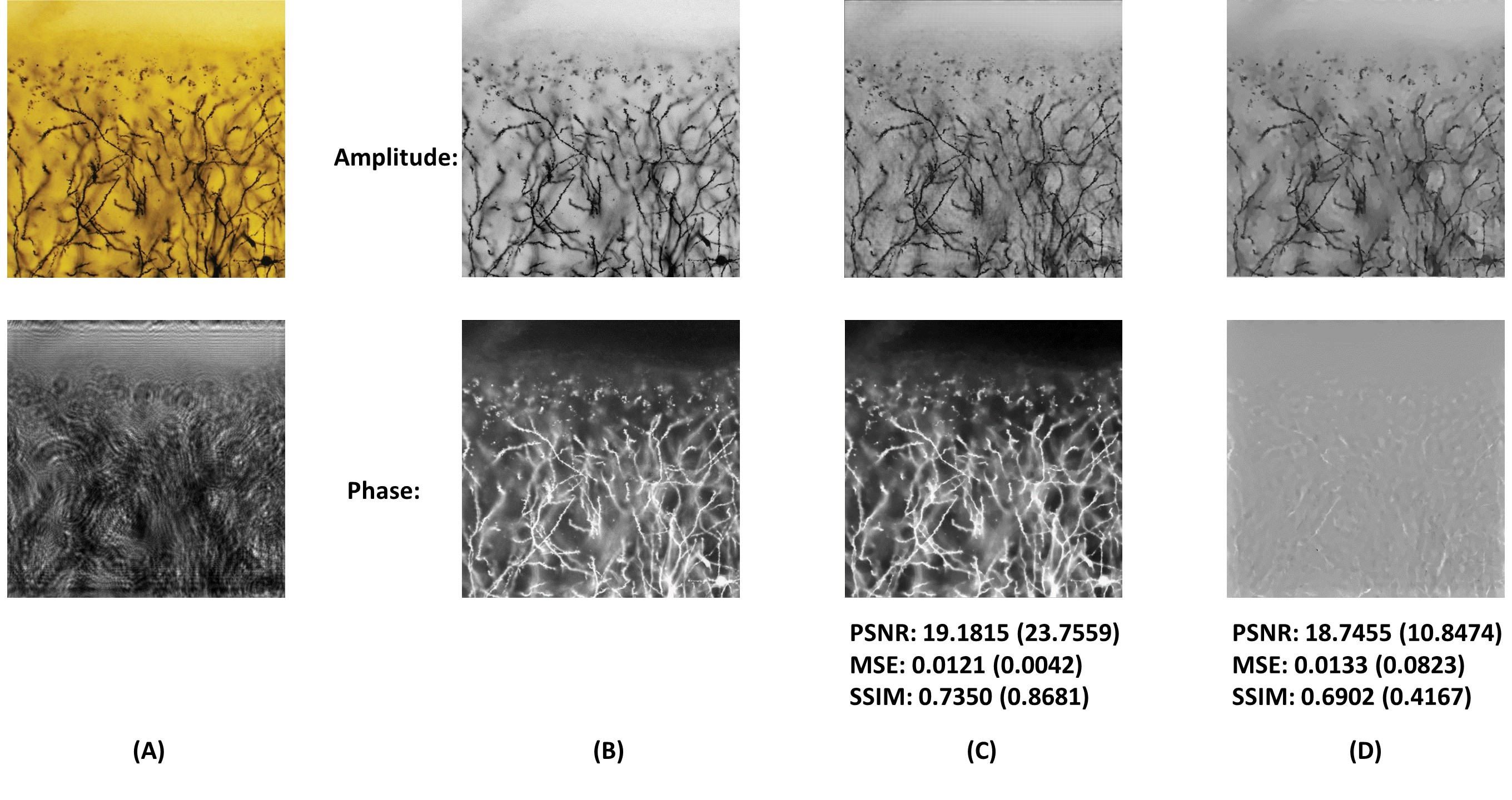}}
\caption{Human dendrite image reconstructions and the evaluating metrics for amplitude (phase): (A) The RGB image and simulated hologram. (B) The reference amplitude and phase. (C) Our method. (D) CS. In this experiment, the proposed method has been shown that have a much better performance on phase information recovery that the CS method.}
\label{ndendrite}
\end{center}
\end{figure}

Further experiments illustrate that the proposed method also dramatically improves the ability to restore detailed textures and phase information. A simulation on cell image shown in Fig.~\ref{fig7} is used to inspect that by taking advantage of the natural superiority of CNNs for image processing problems, the proposed method is able to restore more explicit details on images with more complex structures, as well as more precise phase information. For simulating a hologram with an implicit phase, we apply the grayscale image of the RGB image as the reference amplitude and the green channel as the reference phase. The hologram size is $500\times500$ that are generated with the same light wavelength and object-to-sensor distance on a sensor with a pixel size of 1.67 $\mu m$. Compare with the CS method, our result maintains more structural textures to both the ground truth amplitude and phase. Another simulation followed the same configuration on a human dendrite image to provide a further prof to the outstanding phase reconstruction ability of our method. The reconstruction results are shown in Fig.\ref{ndendrite}.

\begin{figure}[htbp]
\begin{center}
\centerline{\includegraphics[width=0.9\columnwidth]{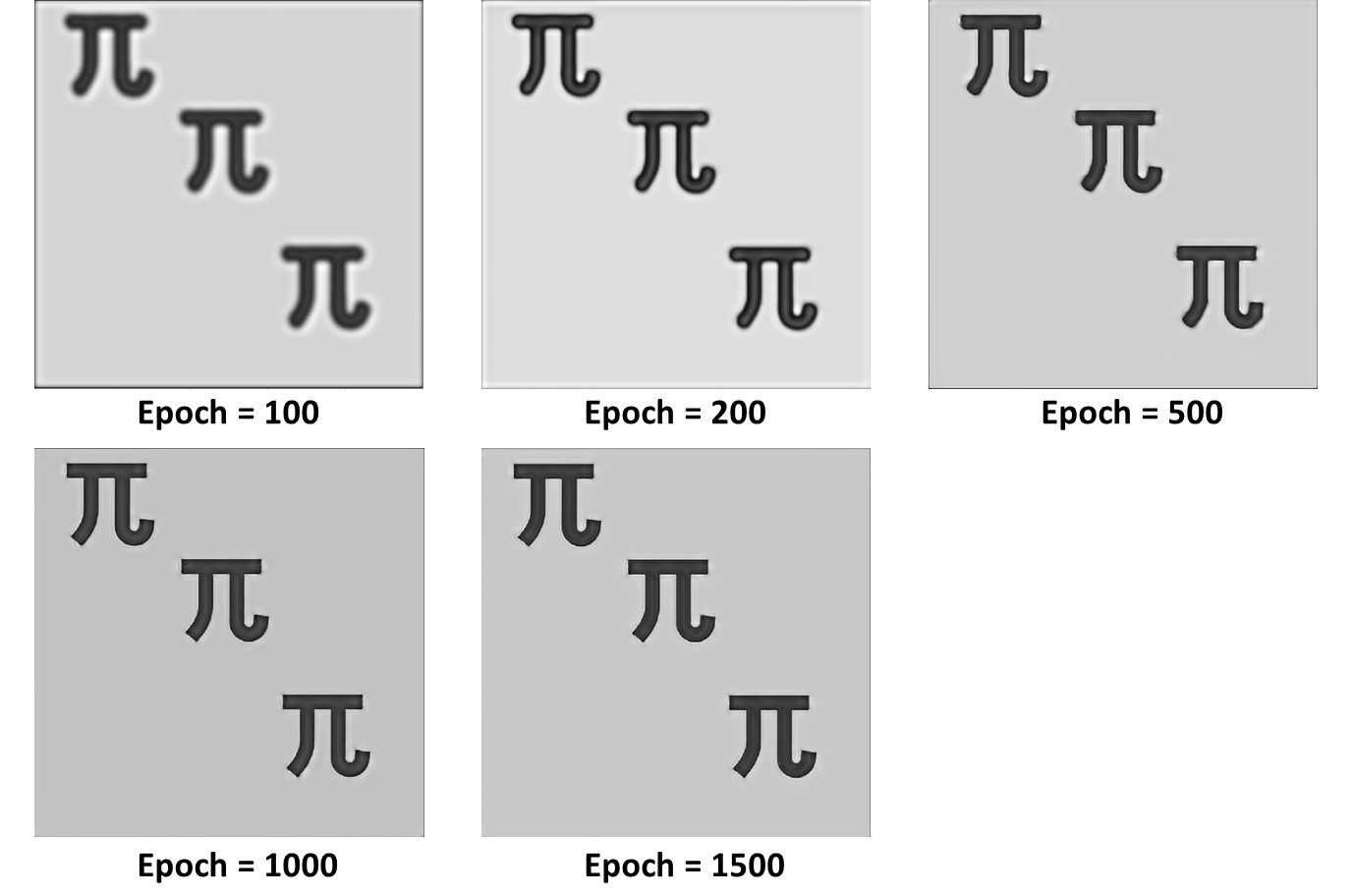}}
\caption{Pi image restoration at 100, 200, 500, 1000, and 1500 training epochs. Obviously, the rough shape of the object is restored first, then more details and sharp edges are restored.}
\label{fig8}
\end{center}
\end{figure}

To reveal the reason why the presented method could fit the networks to obtain the required results, an experiment on a Pi image is conducted. In this experiment, the reconstruction results at different training iterations are shown in Fig.~\ref{fig8}. During the optimization process, the CNNs tend first to restore the general shape of objects and add details to them. This characteristic explains why our method works. Compared with objects, the twin image usually shows a more obscure shape. Therefore, when the network is used to restore the object from a captured hologram, it will converge before the twin image is recovered as the clean object is the solution to the inverse problem.

\begin{figure}[htbp]
\begin{center}
\centerline{\includegraphics[width=0.9\columnwidth]{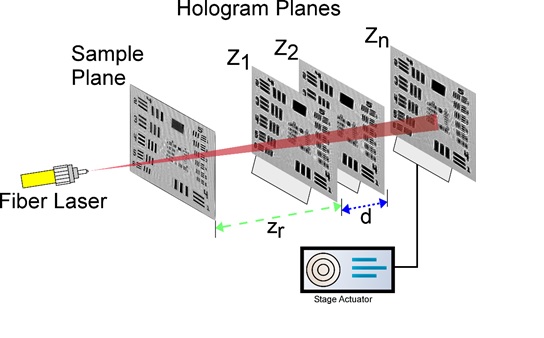}}
\caption{The configuration for the lensless digital Gabor holography system.}
\label{fig9}
\end{center}
\end{figure}

\begin{figure*}[htbp]
\begin{center}
\centerline{\includegraphics[width=1.5\columnwidth]{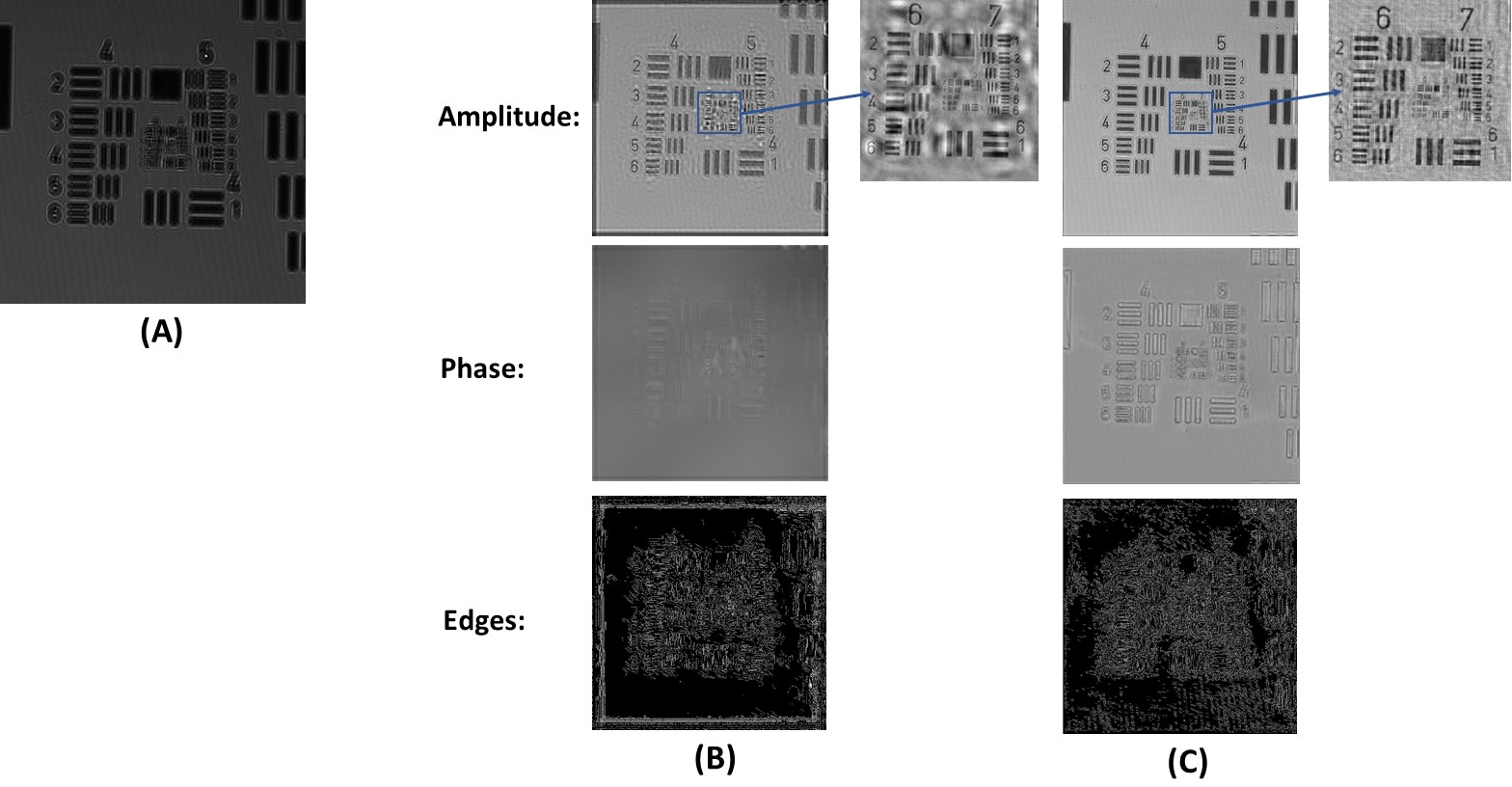}}
\caption{(A) The captured hologram of the USAF positive high-resolution test target. (B) Multi-height reconstruction. (C) The proposed deep learning reconstruction.}
\label{fig10}
\end{center}
\end{figure*}

\begin{figure*}[htbp]
\begin{center}
\centerline{\includegraphics[width=1.5\columnwidth]{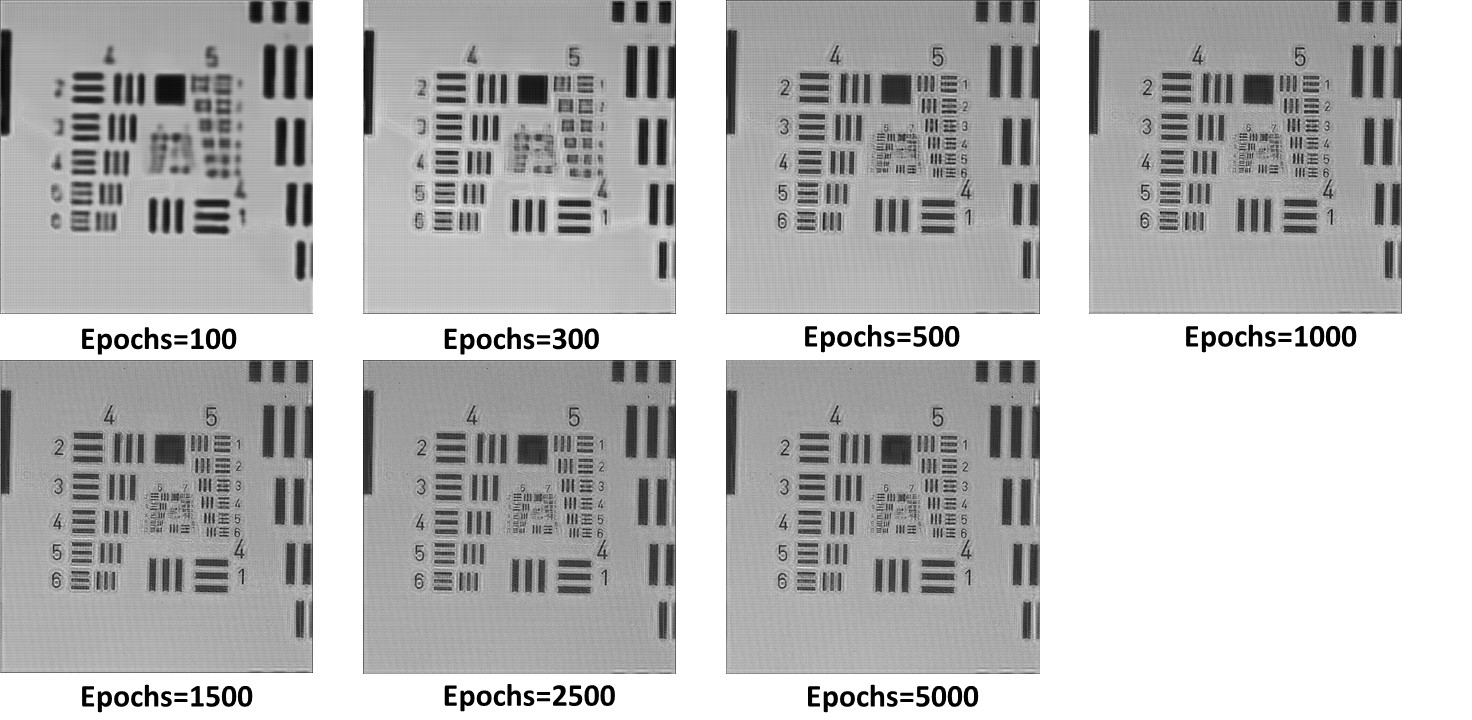}}
\caption{USAF positive high-resolution test target restoration at 100, 300, 500, 1000, 1500, 2500, and 500 training epochs. The reconstruction still follows the regular pattern that the rough shape is restored first and details is restored later.}
\label{fig11}
\end{center}
\end{figure*}

\begin{figure*}[!hbt]
\begin{center}
\centerline{\includegraphics[width=1.5\columnwidth]{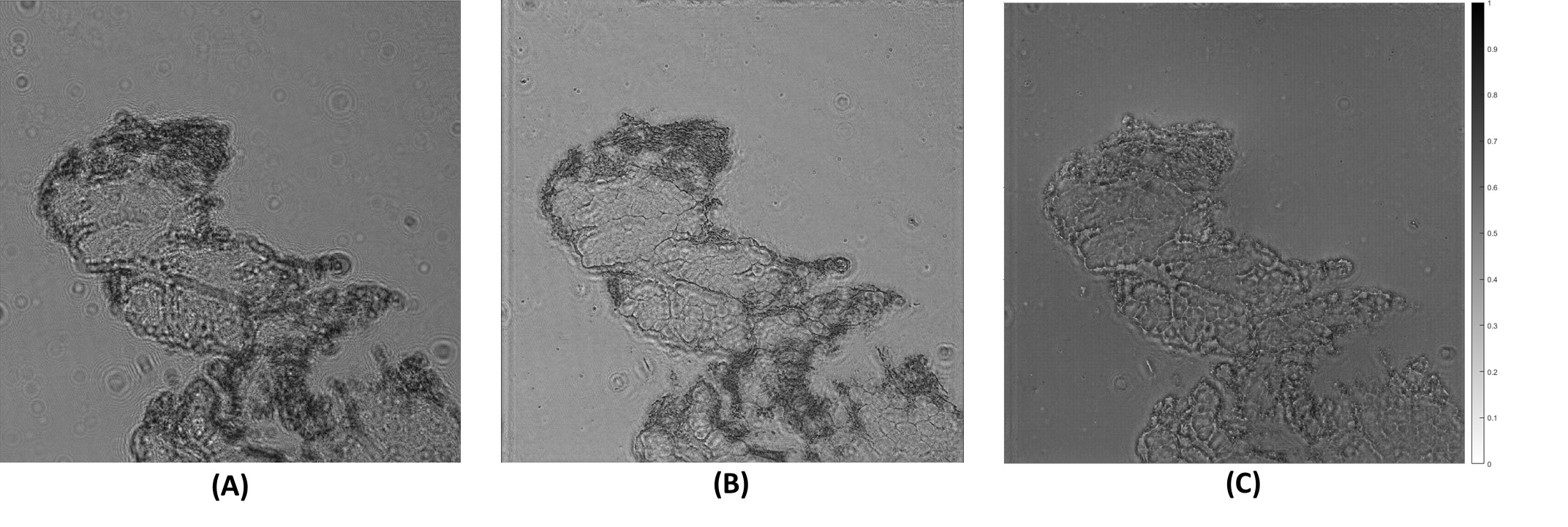}}
\caption{Optical Experimental hologram of USAF Resolution Chart and reconstructions. (A) The captured hologram. (B) Amplitude reconstruction with our method. (C) The reconstructed quantitative phase with our method.}
\label{fig12}
\end{center}
\end{figure*}

\begin{figure*}[!hbt]
\begin{center}
\centerline{\includegraphics[width=1.5\columnwidth]{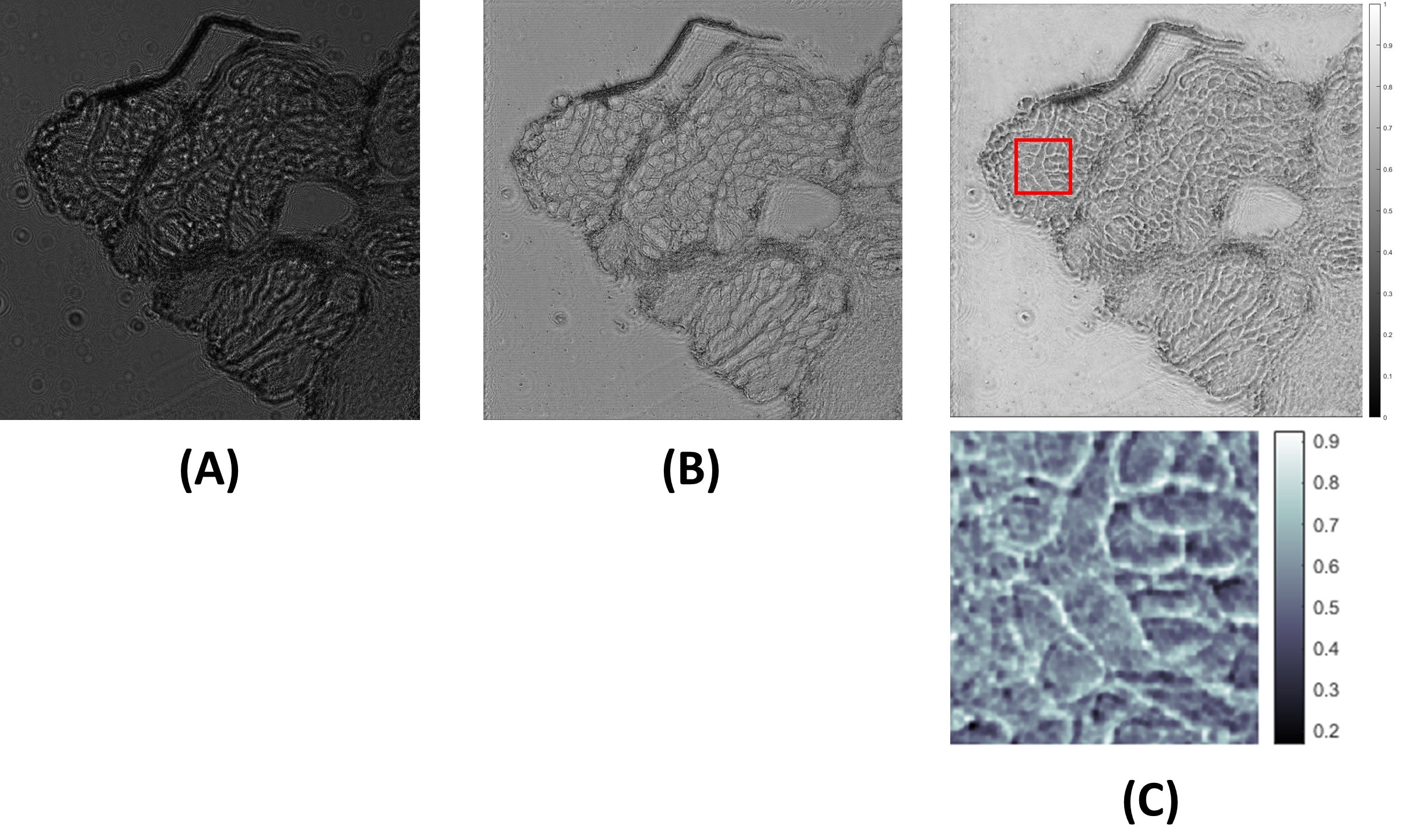}}
\caption{Optical Experimental hologram of a non-keratinizing squamous cell carcinoma and reconstructions. (A) The captured hologram. (B) Amplitude reconstruction with our method. (C) The reconstructed quantitative phase with our method.}
\label{squamous}
\end{center}
\end{figure*}

\section{Optical Experiments}
In order to verify the performance of the proposed method in real-world data, a series of optical experiments are conducted in the laboratory. Fig.~\ref{fig9} illustrates the configuration for the lensless Gabor DHM system used in our experiments. The light source consists of a Thorlabs single mode fiber-coupled laser. A pigtailed light beam is emitted to a single mode fiber that is terminated at an FC/PC bulkhead. The sample is placed between the light source and an image sensor (Imaging Source DMM 27UJ003-ML - pixel size 1.67$\mu m$) with an object reconstruction distance $z$. Performing hologram reconstruction in piratical is a relatively harder task than in simulation as a consequence of the error between the actual parameters and the preset parameters in the experiment. Meanwhile, the influence of ambient light and air dust in the environment leads to high noise in the real hologram. Therefore, the algorithm applied in real-world data is expected to be robust to noise and optical parameter error.

Fig.~\ref{fig10} shows the reconstruction result on a USAF positive high-resolution test target (which means the stripes and digits are thicker than the background). An illuminated plane wave at the wavelength of 406 $\mu m$ is used, and the distance between the target and the image sensor is set at around 1110 $\mu m$. A multi-height TIE based algorithm is used for comparison with ten captured hologram with a step-size $15um$ between the adjacent hologram planes. In previous deep learning based work~\cite{horisaki2018deep,rivenson2018phase,gan2017holography}, multi-height TIE based algorithms are used for producing the ground truth of the training pairs that have been proven to hold an excellent performance. The reconstructed amplitude and phase show the outstanding denoising and twin image removal capability of the proposed method that the reconstructed results have comparable quality to the multi-height methods with single-shot hologram. The enlarged area proves that our approach can retain high-quality details to a great extent while removing the twin image at the same time. The effeteness of the twin image removal ability is quantified as a mean edge factor, which is calculated as $\frac{1}{NM}\sum^{N,M}_{i,j=0}A_{i,j}$, where the $A_{i,j}$ is the edge matrix obtained by the Canny edge detector. We choose the Canny edge detector for getting edge matrix since it is more sensitive than the Sobel operator. The mean edge factors for multi-height method is $0.0990$, respectively, $0.1210$ for our deep learning based methodology.

We also show the reconstruction of our method at different training iterations to examine the theoretical explanation we proposed in Section.~\ref{deep} in Fig.~\ref{fig11}. The results shows that our interpretation of why the presented method works still holds true for real-world data.

An experiment on a sectioned dysplasia tonsillar mucosa tissue is conducted to verify the potential of our method on biomedical usage. The tissue holography could be used to analyze beforehand with clinical histological diagnosis. The hologram is captured with an illuminated plane wave with a wavelength at 0.635 $\mu m$ and an object to sensor distance set at 857 $\mu m$. Fig.~\ref{fig12} shows the captured hologram and reconstruction. The reconstructed phase shows the relative depth of the tissue structure that could be used to reconstruct the 3D surface of the tissue. Another experiment on a non-keratinizing squamous cell carcinoma is shown in Fig.\ref{squamous} also proves the effeteness of the proposed method on biomedical imaging.

\section{Conclusion}
In summary, a deep learning method for single-shot reconstruction of In-line Digital Holography reconstruction is proposed in this paper. The physical symmetry of the holography lead object image and twin image both can be the solution of the hologram. With a given prior, the Auto-encoder is able to reconstruct the object image. The proposed method has been proven powerful and potential through both simulated and optical hologram experiment. Although deep learning based method a relatively time consuming, compared to the complex experimental setup of multi-height phase retrieval, our method is cost-effective.

\bibliography{main}
\bibliographystyle{IEEEtran}
\vspace{12pt}

\end{document}